\documentclass[aps, prl, notitlepage, superscriptaddress, twocolumn, 10pt, floatfix]{revtex4-2}
\usepackage[pdftex,plainpages=false,colorlinks=true,citecolor=blue,linkcolor=blue,urlcolor=blue,filecolor=green,bookmarksopen=true]{hyperref}
\usepackage{amsmath}
\usepackage[caption = false]{subfig}
\usepackage{graphicx,epstopdf}
\usepackage[english]{babel}
\usepackage{blindtext}
\usepackage{lipsum}
\usepackage{amsfonts}
\usepackage{amssymb}
\usepackage{bbm}
\usepackage{enumerate}
\usepackage{color}
\usepackage{latexsym}
\usepackage{times,txfonts}
\usepackage{physics}

\newcommand{\us}{\uparrow}
\newcommand{\ds}{\downarrow}

\DeclareSymbolFont{CMletters}{OML}{cmm}{m}{it}
\DeclareMathSymbol{J}{\mathalpha}{CMletters}{`J}
\DeclareMathSymbol{j}{\mathalpha}{CMletters}{`j}
\DeclareMathSymbol{U}{\mathalpha}{CMletters}{`U}

\begin{document}

\title{Quantum Wheatstone Bridge}

\author{Kasper Poulsen}
\email{poulsen@phys.au.dk}
\affiliation{Department of Physics and Astronomy, Aarhus University, Ny munkegade 120, 8000 Aarhus C, Denmark}

\author{Alan C. Santos}
\email{ac\_santos@df.ufscar.br}
\affiliation{Departamento de F\'{i}sica, Universidade Federal de S\~ao Carlos, Rodovia Washington Lu\'{i}s, km 235 - SP-310, 13565-905 S\~ao Carlos, SP, Brazil}
\affiliation{Department of Physics, Stockholm University, AlbaNova University Center, 106 91 Stockholm, Sweden}

\author{Nikolaj T. Zinner}
\email{zinner@phys.au.dk}
\affiliation{Department of Physics and Astronomy, Aarhus University, Ny munkegade 120, 8000 Aarhus C, Denmark}
\affiliation{Aarhus Institute of Advanced Studies, Aarhus University, Høegh-Guldbergs Gade 6B, 8000 Aarhus C, Denmark}

\begin{abstract}

We propose a quantum Wheatstone bridge as a fully quantum analogue to the classical version. The bridge is a few-body boundary-driven spin chain exploiting quantum effects to gain an enhanced sensitivity to an unknown coupling. The sensitivity is explained by a drop in population of an entangled Bell state due to destructive interference as the controllable coupling approaches the unknown coupling. A simple criteria for the destructive interference is found, and an approximate expression for the width of the drop is derived. The sensitivity to the unknown coupling is quantified using the quantum Fisher information, and we show that the state of the bridge can be measured indirectly through the spin current. Our results are robust towards calibration errors and generic in the sense that several of the current state-of-the-art quantum platforms could be used as a means of realization. The quantum Wheatstone bridge may thus find use in fields such as sensing and metrology using near-term quantum devices.

\end{abstract}

\maketitle

\newpage

\begin{figure}[t]
\centering
\includegraphics[width=1. \linewidth, angle=0]{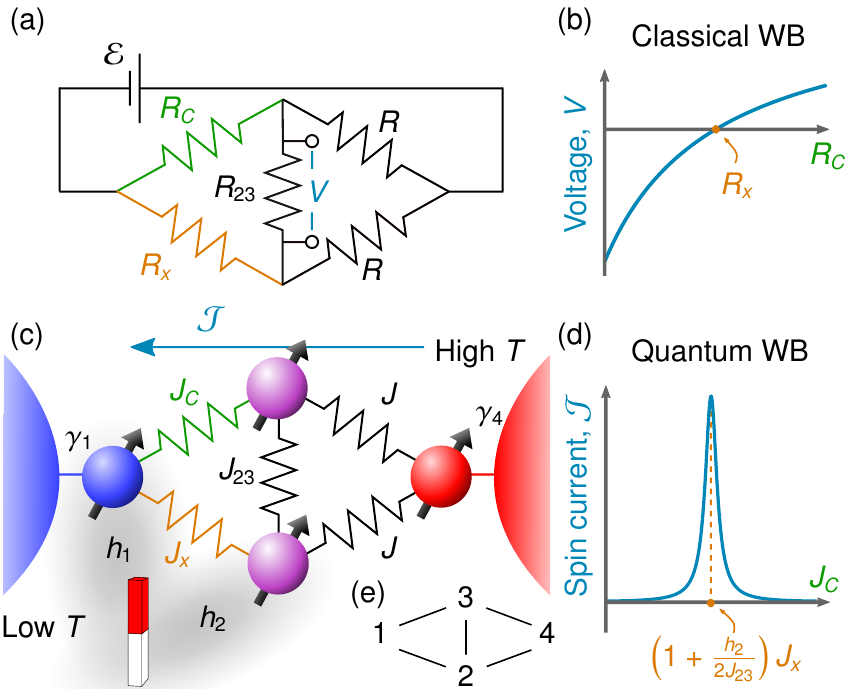}
\caption{(a) Classical Wheatstone bridge consisting of three resistors with known resistances $R$ and $R_C$ and one resistor with unknown resistance $R_x$. (b) Sketch of the voltage dependence used in the classical Wheatstone bridge.
(c) Quantum Wheatstone bridge consisting of four spins interacting with three known coupling strengths $J_C, J$, and $J_{23}$ and one unknown coupling strength $J_x$. Spins 1 and 4 is interacting with two thermal baths of different temperature, and two magnetic fields of strength $h_1$ and $h_2$ is applied to spins 1 and 2, respectively. (d) Sketch of the spin current dependence used in the Quantum Wheatstone bridge.
(e) Numbering for the 4 spins used throughout the text.
}
\label{figure1}
\end{figure}

The classical Wheatstone bridge was first proposed in 1833 by S. H. Christie \cite{Christie1833}, and it was later generalised and popularized by C. Wheatstone \cite{doi:10.1098/rstl.1843.0014}. The bridge, as seen in Fig.~\ref{figure1}(a), is a device used for precise determination of an unknown resistance, $R_x$. In a circuit of four resistances, we have two known resistances (both with resistance $R$), a tunable resistance, $R_C$, and the resistance $R_x$. By varying the controllable resistance until
the voltage over $R_{23}$ is zero (balance point), it is possible to find the value of $R_x$ from the previous knowledge of $R$ and $R_C$. Namely, we get $R_x = R_{C,0}$, where $R_{C,0}$ is the value of the tunable resistance at the balance point.

With increased miniaturization, precise measurement of components remains a challenge. High precision is especially important as we push towards better quantum devices, where accurate knowledge of system parameters are essential for high quality quantum gates \cite{Devitt_2013}. The added control of mesoscopic devices offer new possibilities for precise measurement \cite{Lee2020, PhysRevApplied.10.054048}, and quantum mechanical effects such as entanglement and interference has even been shown to enhance the precision of measurements in certain situations \cite{Pedrozo-Penafiel2020, 4277368, RevModPhys.81.1051}. Recently, it has become possible to control and exploit heat transport through quantum systems.  Of particular interest is boundary driven quantum systems consisting of a quantum system driven away from equilibrium by two thermal baths at the extremities \cite{landi2021non, PhysRevB.96.104402, PhysRevLett.106.217206, PhysRevLett.126.077203}. This facilitates new components like minimal engines \cite{PhysRevE.76.031105, PhysRevLett.125.166802, PhysRevLett.126.120605}, autonomous entanglement engines \cite{Bohr_Brask_2015}, and rectifiers \cite{PhysRevLett.120.200603, PhysRevLett.102.095503, PhysRevE.103.052143, poulsen2021entanglement}. Additionally, the irreversibility of open systems can be used for better measurement \cite{PhysRevX.10.021038, Inomata2016}.

In quantum systems, transport depends on transmission matrix elements between input and output channels, and in turn, this governs the resistance. For spin systems, the matrix elements are parametrized by coupling strengths that serve as the quantum analogue of classical resistances. Here, we introduce a quantum version of the Wheatstone bridge which exploits quantum mechanical effects to gain a large sensitivity to the unknown coupling strength $J_x$. Similar to the original bridge, a controllable coupling, $J_C$, is varied until a certain criteria is met (the balance point). However, here the criteria is the rapid switching of the system state due to perfect destructive interference. The setup itself is comprised of four spins interacting as seen in Fig.~\ref{figure1}(c). Two thermal baths at each extremity create a temperature imbalance thus driving a spin current. The state and the spin current become sensitive to the unknown parameter, and they can both be used for estimation.

The small size of the setup makes it realistic for implementation in the near future using several of the current quantum technology platforms. The quantum Wheatstone bridge is nominally identical to a recent four qubit setup in germanium quantum dots \cite{Hendrickx2021} while superconducting circuits is a proven platform for studying quantum thermodynamics \cite{Ronzani2018, Senior2020}. A possible implementation in superconducting circuits can be found in Supplemental Material.





\noindent {\it Setup.} The model studied is two spins connected through a double-spin interface in a diamond configuration similar to the classical Wheatstone bridge as seen in Fig~\ref{figure1}. The Hamiltonian for the four spins is
\begin{equation}
\begin{aligned}
\hat{H} &= \frac{\omega + 2h_1}{2} \hat{\sigma }_z^{(1)} +  \frac{\omega + 2h_2}{2} \hat{\sigma }_z^{(2)} +  \frac{\omega}{2} \hat{\sigma }_z^{(3)} +  \frac{\omega}{2} \hat{\sigma }_z^{(4)} \\ &\hspace{1.5cm} +  J_x \hat{X}_{12} +  J_C \hat{X}_{13} + J_{23} \hat{X}_{23} + J\hat{X}_{24} + J\hat{X}_{34}, 
\end{aligned}
\end{equation}
where $\hat{X}_{ij} = \hat{\sigma}_x^{(i)} \hat{\sigma}_x^{(j)} + \hat{\sigma}_y^{(i)} \hat{\sigma}_y^{(j)}$ 
is the $XX$ spin exchange operator. The Pauli matrices for the $i$th spin are denoted $\hat{\sigma}_\alpha^{(i)}$ for $\alpha = x,y,z$, and
we are using units where $\hbar=k_B =1$. The parameter $J_x$ is unknown, and $J_C$ is a controllable coupling. 
The angular frequency $\omega$ corresponds to a homogeneous magnetic field, and $h_n$ is an offset in the magnetic field acting on the $n$th spin. The exchange coupling $J$ gives the overall scale of the problem while the exchange between the interface spins is $J_{23}$. 
Here, we focus on the regime $\omega \gg J_{23}, h_1 \gg h_2,J$ and $J_{23} \simeq h_1$. Spin 1 is coupled to a Markovian thermal bath at very low temperature $T_1 \ll \omega + 2h_1$ forcing spin 1 to decay. Similarly, spin 4 is coupled to a Markovian thermal bath of higher temperature $T_4 \sim \omega$ inducing both decay and excitation of spin 4.
The evolution of the system is described using the density matrix $\hat{\rho}$ through the Lindblad master equation \cite{Lindblad1976, breuer2002theory}
\begin{equation}
\frac{d \hat{\rho}}{d t} = -i [\hat{H}, \hat{\rho}] + \mathcal{D}_1[\hat{\rho}] + \mathcal{D}_4[\hat{\rho}]. \label{me:1}
\end{equation}
The Markovian baths are modeled using the non-unitary parts
\begin{equation}
\begin{aligned}
\mathcal{D}_{1}[\hat{\rho}] &= \gamma_1 \mathcal{M}[\hat{\sigma}_-^{(1)}],\\
\mathcal{D}_{4}[\hat{\rho}] &= \gamma_4 (n + 1)\mathcal{M}[\hat{\sigma}_-^{(4)}] + \gamma_4 n \mathcal{M}[\hat{\sigma}_+^{(4)}], \nonumber
\end{aligned}
\end{equation}
where $\mathcal{M}[\hat{A}] = \hat{A} \hat{\rho} \hat{A}^\dag - \frac{1}{2} \{ \hat{A}^\dag \hat{A}, \hat{\rho} \}$. The coupling strength between the cold (hot) bath and spin 1 (spin 4) is $\gamma_1$ ($\gamma_4$). The mean number of excitations in the hot bath mode of energy $\omega$ is $n = \big( e^{\omega/T_4} -1 \big)^{-1}$, where $T_4$ is the temperature of the hot bath. Due to the baths, the quantum Wheatstone bridge is insensitive to decoherence on spin 1 and spin 4. The two baths will induce heat flow and generally drive the system into a non-equilibrium state. After sufficient time, the system will reach a steady state, $\partial \hat{\rho}_\text{ss}/\partial t = 0$. It is this steady state that we will use to probe the value of $J_x$. Unless otherwise stated, throughout our work we will use the set of parameters $J_{23} = 20J, h_1 = 20J, h_2 = 0.5J, J_x = J, n = 0.5, \gamma_1 = J, \gamma_4 = 10J$.
The excitation energy $\omega$ has no impact on the steady state and is not set. This is seen by transforming the Lindblad master equation into the interaction picture with respect to $\hat{H}_0 = \frac{\omega}{2} \sum_{i=1}^4 \hat{\sigma}_z^{(i)}$. Since the Hamiltonian is spin preserving, $\hat{\rho}_{\text{ss}}$ is unchanged.

\begin{figure}[t]
\centering
\includegraphics[width=1. \linewidth, angle=0]{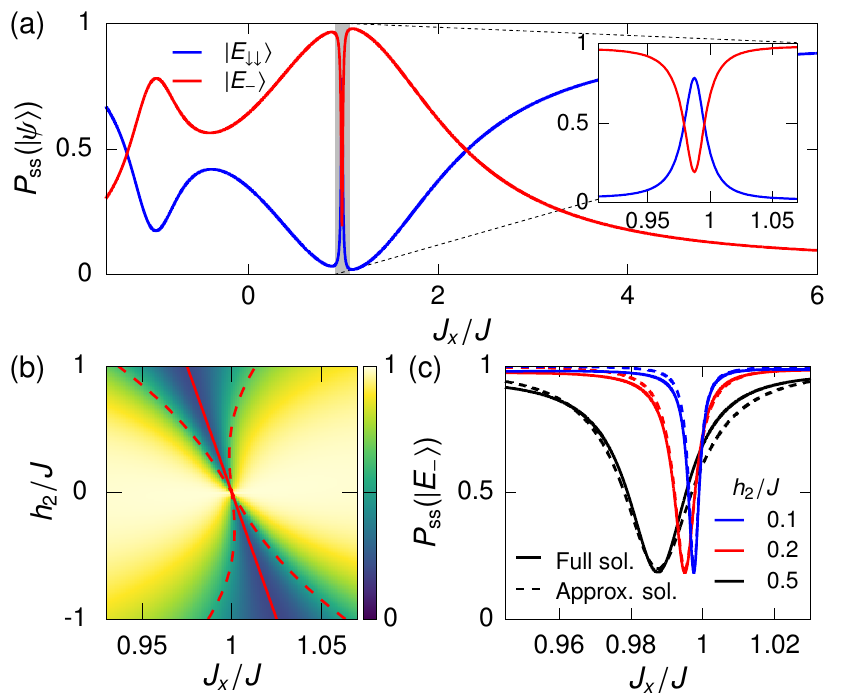}
\caption{(a) Population of the two states $\ket*{E_{\ds \ds}}$ and $\ket{E_-}$ as a function of the unknown parameter $J_x$. (b) Population of $\ket{E_-}$ as a function of both $J_x$ and $h_2$. The solid red line shows the points of destructive interference, Eq.~\eqref{resonance}, while the dashed lines denote the width of the population drop, Eq.~\eqref{width}. 
(c) Population of $\ket{E_-}$ as a function of $J_x$ for both the full numerical solution (solid line) and the approximate solution (dashed line) given by Eq.~\eqref{pop1}. Here $J_C=J$ was used.
} 
\label{figure2}
\end{figure}

{\it Sensitivity to $J_x$}. The steady state of the two interface spins is sensitive to changes in the unknown parameter $J_x$. This sensitivity can be used to determine the value of $J_x$. Because of the two baths, the coherences between the two interface spins and the outer spins are negligible, and the density matrix is well described by the populations. Therefore, the Hamiltonian for spins 2 and 3 is diagonalized and the eigenstates, to linear order in $h_2/J_{23}$, are found to be
\begin{subequations}
\label{states}
\begin{alignat}{1}
\ket*{E_{\us \us}} = \ket{\us \us},\quad
&\ket{E_+} = \ket{\Psi_+} + \frac{h_2}{4J_{23}} \ket{\Psi_-},\\
\ket*{E_{\ds \ds}} = \ket{\ds \ds},\quad &\ket{E_-} = \ket{\Psi_-} - \frac{h_2}{4J_{23}} \ket{\Psi_+} ,
\end{alignat}
\end{subequations}
where $\ket{\Psi_\pm} = (\ket{\us \ds} \pm \ket{\ds \us})/\sqrt{2}$. 
For the reduced density matrix $\hat{\rho}_{\text{ss}}^{(23)} = \tr_{(1,4)} \{\hat{\rho}_{\text{ss}}\}$, the populations for the two states $\ket*{E_{\ds \ds}}$ and $\ket{E_-}$ are shown in Fig.~\ref{figure2}(a) as a function of $J_x$. 
The populations are given by $P_{\text{ss}}(\ket*{E_{\downarrow \downarrow}}) = \tr_{(2,3)} \{\hat{\rho}_{\text{ss}}^{(23)} \op*{E_{\downarrow \downarrow}}\}$ and $P_{\text{ss}}(\ket*{E_{-}}) = \tr_{(2,3)} \{\hat{\rho}_{\text{ss}}^{(23)} \op*{E_{-}}\}$, where $\text{tr}_{(2,3)}\{ \bullet \}$ is the trace of the Hilbert space for spins 2 and 3. This shows that the two interface spins are predominantly in these two states.
The two spins are driven into the entangled state $\ket{E_-}$ as $J_x$ approaches $J_C$ \cite{poulsen2021entanglement}. This can be explained through the transitions between the states of spins 1, 2, and 3 given by
\begin{equation}
\ket{\ds \us \us} \leftrightarrow \ket{\us E_-} \rightarrow \ket{\ds E_-}. \label{transition}
\end{equation}
The parameters of the model are chosen such that the transition $\ket{\ds \us \us} \leftrightarrow \ket{\us E_-}$ is energetically allowed, and the transition $\ket{\us E_-} \rightarrow \ket{\ds E_-}$ makes the process irreversible. The present scheme is very different from \cite{poulsen2021entanglement}, as the population of $\ket{E_-}$ drops suddenly, which can be seen in the inset of Fig.~\ref{figure2}(a). 
This sudden change in the populations is what can be exploited for precise determination of $J_x$.

In Fig. \ref{figure2}(b), the population of $\ket{E_-}$ is plotted as a function of both the magnetic field $h_2$ and the unknown parameter $J_x$. It is seen that the value of $J_x$ for which the population is minimal is linear in $h_2$. This can be explained by looking at the matrix element for the first transition in Eq.~\eqref{transition} to linear order in $h_2/J_{23}$
\begin{equation}
\begin{aligned}
&\mel{\us E_- \us}{\hat{H}}{\ds \us \us \us} = \mel{\us E_- \ds}{\hat{H}}{\ds \us \us \ds} \\
&\hspace{2cm}= -\sqrt{2} \left[ J_x \left( 1+\frac{h_2}{4J_{23}} \right) - J_C \left( 1-\frac{h_2}{4J_{23}} \right) \right]. 
\end{aligned}
\end{equation}
The transition is forbidden to first order when the above matrix element is zero. This is equivalent to destructive interference occurring between the spin excitation located at spin 2 and the spin excitation located at spin 3. The destructive interference condition is $J_x = J_{x,0}$, where
\begin{equation}
J_{x,0} = J_C \left( 1- \frac{h_2}{2J_{23}} \right). \label{resonance} 
\end{equation}
This parametrization is plotted as a solid red line in Fig. \ref{figure2}(b). To analytically approximate the width of the dip in population shown in Fig.~\ref{figure2}(a), we use the master equation for the four states of the interface spins
\begin{equation}
\dot{\vec{P}} = \mathbf{W} \vec{P},
\end{equation}
where $\vec{P} = [P(\ket*{E_{\ds \ds}}), P(\ket{E_{-}}), P(\ket{E_{+}}), P(\ket*{E_{\us \us}})]^T$ and $\mathbf{W}$ is a matrix of rates, $\Gamma_{\ket{E}\rightarrow \ket{E'}}$, describing the rate of transfer from $\ket{E}$ to $\ket{E'}$. The rates are found around the point $J_x = J_{x,0}$
\begin{subequations}
\label{rates}
\begin{alignat}{1}
\Gamma_{\ket*{E_{\us \us}} \rightarrow \ket{E_-}} &\simeq \frac{2 \gamma_1 \left(J_x - J_{x,0} \right)^2}{ (2h_1 - 2J_{23} - h_2)^2 + \gamma_1^2/4 }, \\
\Gamma_{\ket{E_-} \rightarrow \ket*{E_{\ds \ds}}} &\simeq \frac{ h_2^2 J^2 (n+1) \gamma_4}{2J_{23}^2 \eta^2},\,
\Gamma_{\ket*{E_{\us \us}} \rightarrow \ket{E_+}} \simeq \frac{ 8J^2 (n+1) \gamma_4}{\eta^2},  \\
\Gamma_{\ket*{E_{\ds \ds}} \rightarrow \ket{E_{-}}} &\simeq \Gamma_{\ket*{E_-} \rightarrow \ket{E_{\us \us}}} \simeq  \frac{h_2^2 J^2 n \gamma_4 }{2 J_{23}^2 \eta^2},\, 
\Gamma_{\ket*{E_{+}} \rightarrow \ket*{E_{\ds \ds}}} \simeq \gamma_1, \\
\Gamma_{\ket*{E_{\ds \ds}} \rightarrow \ket{E_{+}}} &\simeq \Gamma_{\ket*{E_+} \rightarrow \ket{E_{\us \us}}} \simeq \frac{ 8J^2 n \gamma_4}{\eta^2},
\end{alignat}
\end{subequations}
where $\eta^2 = 4J_{23}^2 + (2n+1)^2\gamma_4^2 /4$. All other rates are zero due to spin conservation, and the full list of assumptions to arrive at these rates is: $J\sim J_C \sim J_x$; $h_2,J \ll J_{23}$; $h_1 \simeq J_{23}$; $2\left(J_x - J_{x,0} \right)^2 \ll (2h_1 - 2J_{23} - h_2)^2 + \gamma_1^2/4 \ll 8J_C^2$; $\gamma_1 \ll \gamma_4 \leq 4 J_{23}/(2n+1)$; $n\gg J/J_{23}$; and $|J_x-J_{x,0}| \ll J$. See Supplemental Material for the full form of $\mathbf{W}$ and further discussion of the assumptions. Solving for steady state, $\mathbf{W} \vec{P}_{\text{ss}} = 0$, the population of $\ket{E_-}$ takes the form
\begin{equation}
P_{\text{ss}}(\ket{E_-}) \simeq \frac{\left(J_x - J_{x,0} \right)^2 + P_{\text{ss},-}^0 \frac{\Lambda^2}{4}}{\left(J_x - J_{x,0} \right)^2 + \frac{\Lambda^2}{4}}. \label{pop1}
\end{equation}
$P_{\text{ss},-}^0 = \frac{n}{3n+1}$ is the population at $J_x = J_{x,0}$, and $\Lambda$ is the width of the Lorenzian,
\begin{equation}
\Lambda = \sqrt{\frac{(n+1)(3n+1)}{2n^2}} \frac{ h_2  \sqrt{\left[2h_1 - 2J_{23} - h_2\right]^2 + \gamma_1^2/4 }  }{2 J_{23} }. \label{width}
\end{equation}
The dashed red lines in Fig.~\ref{figure2}(b) correspond to the two lines $J_x = J_{x,0} \pm \Lambda /2$. 
To further explore the validity of the approximate expression for $P_{\text{ss}}(\ket{E_-})$, we plot both the exact numerical solution to the Lindblad master equation and the approximate solution in Fig~\ref{figure2}(c). The two solutions have small deviations which become greater for larger $h_2$ as expected from the assumptions. Overall, there is good agreement between the two solutions justifying the assumptions.
Furthermore, the width, and thus the sensitivity to $J_x$, can be tuned through the ratio $h_2/J_{23}$ in agreement with Eq.~\eqref{width}. This is particularly useful for calibration of $h_2$, since $h_2\simeq 0$ can be found by minimizing the width of the population with respect to $h_2$.

\begin{figure}[t]
\centering
\includegraphics[width=1. \linewidth, angle=0]{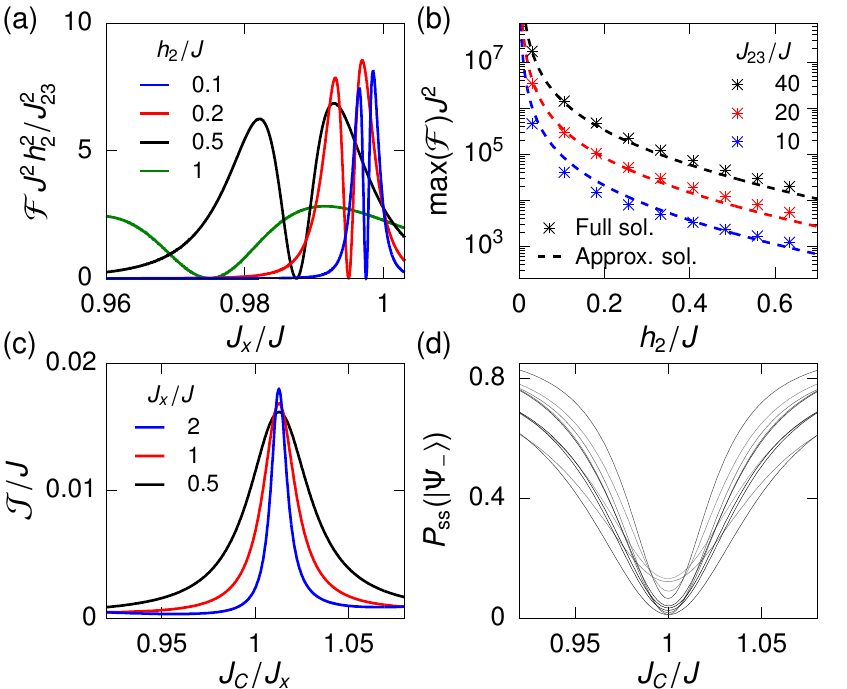}
\caption{(a) Quantum Fisher information, $\mathcal{F}$, as a function of the unknown parameter $J_x$ for different values of the magnetic field $h_2$ and $J_C = J$. (b) Maximum Quantum Fisher information for both the full numerical solution and the approximate solution given in Eq.~\eqref{max_FI} and using $J_C = J$. (c) Spin current, $\mathcal{J}$, as a function of $J_C$ for different values of the unknown parameter $J_x$. (d) $P_\text{ss}(\ket{\Psi_-})$ as a function of $J_C$ for 10 sets of random errors on all parameters except $h_2$ and $J_x$ (see text) where $h_2 = 0$.  
}
\label{figure3}
\end{figure}

{\it Quantum Fisher information.} A measure of the sensitivity of the density matrix to small variations in the unknown parameter $J_x$ is the quantum Fisher information \cite{PhysRevA.97.042322, CHAPEAUBLONDEAU20171369}, which for a diagonal density matrix, $\hat{\rho}_{\text{ss}} = \sum_k p_{k} \op{k}$, is
\begin{equation}
\mathcal{F} = 2\sum_{p_k+p_l > 0} \frac{\mel{k}{\partial_{J_x} \hat{\rho}_{\text{ss}}}{l} \mel{l}{\partial_{J_x} \hat{\rho}_{\text{ss}}}{k}}{p_k + p_l}.
\end{equation}
The ultimate limit in precision is then given by the Cramér-Rao bound $\text{Var}(J_x) \geq 1/\mathcal{F}$ for a single shot measurement. The quantum Fisher information is plotted in Fig.~\ref{figure3}(a) for different values of $h_2$. Comparing Fig.~\ref{figure2}(c) and Fig.~\ref{figure3}(a), we see that the largest quantum Fisher information overlaps with the largest change in the populations. 
Furthermore, we see that the maximum quantum Fisher information is $\text{max}(\mathcal{F}) \propto J_{23}^2/h_2^2$. 
To explain this, we note that the populations $P_{\text{ss}}(\ket*{E_+})$ and $P_{\text{ss}}(\ket*{E_{\us \us}})$ are of order $J^2/J_{23}^2$. 
Therefore, the quantum Fisher information can be found using the population in Eq.~\eqref{pop1} and $P_{\text{ss}}(\ket*{E_{\ds \ds}}) \simeq 1-P_{\text{ss}}(\ket*{E_{-}})$,
\begin{equation}
\mathcal{F} \simeq \frac{4 \left(J_x - J_{x,0}\right)^2 \Lambda^2 \left(2 n + 1 \right)}{\left(\left(J_x - J_{x,0}\right)^2 + \frac{\Lambda^2}{4}\right)^2 \left(n\Lambda^2+4(3n + 1)\left(J_x - J_{x,0}\right)^2 \right)}.
\end{equation}
The maximum quantum Fisher information is found by maximizing the above expression
\begin{equation}
\text{max}(\mathcal{F}) \simeq \frac{4 N(n)}{\Lambda^2}, \label{max_FI}
\end{equation}
where $\frac{3}{4} < N(n) \leq 4$ is a function of $n$ only, and $N(n=0.5) \simeq 1.14$, see Supplemental Material for more information. In Fig.~\ref{figure3}(b), the maximum quantum Fisher information is plotted for both the full numerical solution and the approximate expression above. The full numerical solution is found by optimizing the quantum Fisher information with respect to $J_x$ around $J_x \sim J_C$. We observe that the agreement is generally good, and that it is better in the expected limit $h_2, J \ll J_{23}$.

{\it Measuring the interface state.} The general operation of the quantum Wheatstone bridge is to slowly vary $J_C$ until the population of $\ket{E_-}$ drops.
The unknown parameter $J_x$ is then either determined by finding the minimum population at the balance point and subsequently using the relation \eqref{resonance}, or it is determined to greater precision by using the slope of the population thus exploiting the full potential found through the quantum Fisher information. In the Supplemental Material, we show that a measurement of the state $\ket{E_-}$ saturates the quantum Fisher information, and we provide a way of performing such a measurement through a shadow spin. Alternatively, the states $\ket*{E_{\ds \ds}}$ and $\ket{E_-}$ can be distinguished by measuring $\hat{\sigma}_z$ for either spin 2 or spin 3. If the measurement yields $-1$ every time, the state is $\ket*{E_{\ds \ds}}$, and otherwise, it is $\ket{E_-}$.
However, such a measurement will destroy the state $\ket{E_-}$, and it does not saturate the quantum Fisher information, see Supplemental Material.
Another alternative is to probe the interface state through the spin current. The spin current is defined as the number of spin excitations decaying to the cold bath per unit of time
$\mathcal{J} = \gamma_1 \mathrm{tr} \{\hat{\sigma}_+^{(1)} \hat{\sigma}_-^{(1)} \hat{\rho}_\mathrm{ss} \}$ \cite{PhysRevE.97.022115}, where $\text{tr}\{\bullet\}$ is the trace over the entire Hilbert space.
The heat, $\mathcal{K}$, exchanged between the system and the cold bath is equal to the spin current times the energy of a single spin excitation $\mathcal{K} \simeq \omega \mathcal{J}$ for $\omega \gg J_{23},J$. Therefore, $\mathcal{K}$ offers an alternative way of measuring $\mathcal{J}$. The entangled state results in a small current due to interference while the state $\ket*{E_{\ds \ds}}$ allows for a current through $\ket{E_+}$. 
In Fig. \ref{figure3}(c), the current is plotted as a function of $J_C$ for different values of $J_x$. First, we notice that the current peaks at the same value of the ratio $J_C/J_x$. 
Second, we notice that the current peaks at the expected $J_C = J_{C,0}$ where
\begin{equation}
J_{C,0} = J_x \left( 1+ \frac{h_2}{2J_{23}} \right). 
\end{equation}
As mentioned, this shows that the spin current can be used to probe $J_x$ without performing measurements on spins 2 and 3 directly. Therefore, the balance point for the quantum Wheatstone bridge is $J_C = J_{C,0}$.

In order to describe the width of the peak in current, we can again solve the steady state, $\mathbf{W} \vec{P}_{\textbf{ss}}=0$, for the rates in Eq.~\eqref{rates}. The solution is found as a function of $J_C$ for $|J_{C,0} - J_C| \ll J$, reflecting the fact that the operation of the quantum Wheatstone bridge involves varying $J_C$ in order to determine $J_x$. 
The current is dominated by the decay of $\ket*{E_{+}}$, and it takes the simplified form $\mathcal{J} \simeq \gamma_1 P_{\text{ss}}(\ket{E_+})$. The approximate expression for the current is
\begin{equation}
\mathcal{J} \simeq \frac{\mathcal{J}_{\infty} (J_C - J_{C,0})^2 + \mathcal{J}_{0} \frac{\Lambda^2}{4}}{(J_C - J_{C,0})^2 + \frac{\Lambda^2}{4}}.
\end{equation}
$\mathcal{J}_0 =8 n (2n+1 ) \gamma_4 J^2/( 3n+1) \eta^2$ is the current at $J_C = J_{C,0}$, and $\mathcal{J}_{\infty} = (n+1) \gamma_1 h_2^2/16 n J_{23}^2$ is the current for $|J_{C,0}-J_C| \rightarrow \infty$. Note that the expression is not valid for $|J_{C,0}-J_C| \rightarrow \infty$, and $\mathcal{J}_\infty$ should not be taken literally. The width of the current is the same as for $P_{\text{ss}} (\ket*{E_-})$ given in Eq.~\eqref{width}. Importantly, the current peaks at the same value $J_C = J_{C,0}$ as the population $P_{\text{ss}}(\ket{E_-})$ drops. 

Finally, we study the effect of calibration errors for all parameters. $h_2$ is already known to shift and widen the Lorenzian, so we will focus on the other parameters. In Fig.~\ref{figure3}(d), $P_\text{ss}(\ket{\Psi_-})$ is plotted as a function of $J_C$ for random normal distributed errors on all parameters except $h_2$ and $J_x$. The standard deviations are $0.02J$ for couplings, $0.2J$ for magnetic fields, $0.1J$ for bath coupling rates, and $0.1$ for $n$. The errors on the coupling of spin 2 to spin 4 and the coupling of spin 3 to spin 4 where sampled separately. The error on the magnetic field for spins 2 and 3 is the same to keep $h_2=0$, and an error was also added to the magnetic field for spin 4. Furthermore, decoherence was implemented through $\mathcal{L} \rightarrow \mathcal{L} + \mathcal{D}_2 + \mathcal{D}_3$, where $\mathcal{D}_i = T^{-1} \mathcal{M}[\hat{\sigma}_-^{(i)}] + T^{-1} \mathcal{M}[\hat{\sigma}_z^{(i)}/2]$ for $i=2,3$, and $TJ = 4\cdot 10^4$ \cite{devoret2013}. Remarkably, the minimum is at $J_C=J_x$ for all 10 samples, and the calibration errors do not result in systematic errors. Furthermore, $h_2 = 0$ should give an infinitely narrow drop, however, due to decoherence the drop will have finite width even for perfect parameters. Therefore, $h_2 = 0$ can be found be minimizing the width with respect to $h_2$.

{\it Conclusion.} We have proposed a quantum mechanical version of the Wheatstone bridge consisting of two interface qubits coupled to a cold qubit and a hot qubit on each side. The state of the interface qubits exhibit large sensitivity to the unknown parameter $J_x$ which can be used for precise measurement. This was explained by an interference criteria, $J_x =J_{x,0}$, preventing the transition $\ket{\uparrow \uparrow} \rightarrow \ket{E_-}$. The sensitivity is quantified using the quantum Fisher information, and Fisher information values of $>10^5$ can be reached. 
Furthermore, we derived approximate expressions for the populations, Fisher information, and spin current. 
We showed that the spin current is a good observable for finding the balance point of the quantum Wheatstone bridge, $J_C = J_{C,0}$. Finally, the mechanism was shown to be robust towards calibration errors.

The model is a generic model of interacting two level systems with many possible implementations. Of particular note is germanium quantum dots \cite{Hendrickx2021, Lawrie2020} and superconducting circuits \cite{rasmussen2021superconducting, Ronzani2018, Senior2020}. A possible implementation using superconducting circuits is proposed in the Supplemental Material.

\begin{acknowledgments}
K.P. and N.T.Z. acknowledge funding from The Independent Research Fund Denmark DFF-FNU. A.C.S. acknowledges financial support from the São Paulo Research Foundation (FAPESP) (Grant No 2019/22685-1 and 2021/10224-0)
\end{acknowledgments}

\bibliography{bibliography}

\onecolumngrid
{\color{white} 0}

\newpage

\renewcommand{\thesubsection}{Appendix A\arabic{subsection}}
\appendix

\section*{ \huge{ S\lowercase{upplemental} M\lowercase{aterial} }}

\section{Approximate expressions for populations, spin current, and quantum Fisher information.}

The approximate expressions for the populations, spin current, and quantum Fisher information in the main text can be found by considering the eigenstates of the interface Hamiltonian. The Hamiltonian for spins 2 and 3, to linear order in $h_2/J_{23}$, has eigenenergies
\begin{subequations}
\begin{alignat}{1}
E_{\us \us} &= \omega + h_2,\\
E_+ &= 2J_{23},\\
E_- &= -2J_{23},\\
E_{\ds \ds} &= -\omega - h_2,
\end{alignat}
\end{subequations}
with corresponding states $\ket*{E_{\us \us}}, \ket{E_+}, \ket{E_-}$ and $\ket*{E_{\ds \ds}}$. The two baths drive transitions between these four states with rates that are approximately
\begin{subequations}
\begin{alignat}{3}
\Gamma_{\ket*{E} \rightarrow \ket*{E'}}^C &= \frac{\big|M_{\ket*{E} \rightarrow \ket*{E'}}^C\big|^2 \gamma_1}{\big|M_{\ket*{E} \rightarrow \ket*{E'}}^C\big|^2 + (\omega + 2h_1 + E' - E)^2 + \gamma_1^2/4},\quad &  M_{\ket*{E} \rightarrow \ket*{E'}}^C &= 2\mel*{E'}{J_x \hat{\sigma}_-^{(2)} + J_C \hat{\sigma}_-^{(3)}}{E},\\
\Gamma_{\ket*{E'} \rightarrow \ket*{E}}^C &= 0, & &\\
\Gamma_{\ket*{E} \rightarrow \ket*{E'}}^H &= \frac{\big|M_{\ket*{E} \rightarrow \ket*{E'}}^H\big|^2 \gamma_4 (n+1)}{(\omega + E' - E)^2 + \gamma_4^2 (2n+1)^2 /4}, & M_{\ket*{E} \rightarrow \ket*{E'}}^H &= 2J\mel*{E'}{\hat{\sigma}_-^{(2)} + \hat{\sigma}_-^{(3)}}{E},\\
\Gamma_{\ket*{E'} \rightarrow \ket*{E}}^H &= \frac{\big|M_{\ket*{E'} \rightarrow \ket*{E}}^H\big|^2 \gamma_4 n}{(\omega + E - E')^2 + \gamma_4^2 (2n+1)^2 /4}, & M_{\ket*{E'} \rightarrow \ket*{E}}^H &= 2J\mel*{E}{\hat{\sigma}_-^{(2)} + \hat{\sigma}_-^{(3)}}{E'},
\end{alignat}
\end{subequations}
for $(E, E') \in \{(E_{\us \us}, E_-), (E_{\us \us}, E_+), (E_{-}, E_{\ds \ds}), (E_+, E_{\ds \ds})\}$. The total rates are defined by $\Gamma_{\ket{E} \rightarrow \ket{E'}} = \Gamma_{\ket{E} \rightarrow \ket{E'}}^C + \Gamma_{\ket{E} \rightarrow \ket{E'}}^H$. The cold bath rates are found by adding the rate of interaction between the interface and the cold qubit from Fermi's golden rule with the rate of decay for the cold qubit \cite{Kapit_2017}. The hot bath rates are found by using the Markov approximation for the hot qubit, thus assuming that the hot qubit correlation functions decay faster than the coupling to the interface \cite{breuer2002theory}. Writing the populations of the four states in a vector $\vec{P} = [P(\ket*{E_{\ds \ds}}), P(\ket{E_{-}}), P(\ket{E_{+}}), P(\ket*{E_{\us \us}})]^T$, the time evolution can be written
\begin{equation}
\dot{\vec{P}} = \mathbf{W} \vec{P},
\end{equation}
where
\begin{equation}
\mathbf{W} = \begin{pmatrix}
-\Gamma_{\ket*{E_{\ds \ds}} \rightarrow \ket{E_-}}-\Gamma_{\ket*{E_{\ds \ds}} \rightarrow \ket{E_+}} & \Gamma_{\ket*{E_{-}} \rightarrow \ket*{E_{\ds \ds}}} & \Gamma_{\ket*{E_{+}} \rightarrow \ket*{E_{\ds \ds}}} & 0\\
\Gamma_{\ket*{E_{\ds \ds}} \rightarrow \ket{E_-}} & -\Gamma_{\ket*{E_{-}} \rightarrow \ket*{E_{\ds \ds}}} -\Gamma_{\ket*{E_{-}} \rightarrow \ket*{E_{\us \us}}} & 0 & \Gamma_{\ket*{E_{\us \us}} \rightarrow \ket*{E_-}}\\
\Gamma_{\ket*{E_{\ds \ds}} \rightarrow \ket{E_+}} & 0 & -\Gamma_{\ket*{E_{+}} \rightarrow \ket*{E_{\ds \ds}}} - \Gamma_{\ket*{E_{+}} \rightarrow \ket*{E_{\us \us}}} & \Gamma_{\ket*{E_{\us \us}} \rightarrow \ket*{E_+}}\\
0 & \Gamma_{\ket*{E_{-}} \rightarrow \ket*{E_{\us \us}}} & \Gamma_{\ket*{E_{+}} \rightarrow \ket*{E_{\us \us}}} & -\Gamma_{\ket*{E_{\us \us}} \rightarrow \ket*{E_-}}-\Gamma_{\ket*{E_{\us \us}} \rightarrow \ket*{E_+}}
\end{pmatrix}.
\end{equation}

After a sufficiently long time, the populations will reach a steady state for which $\mathbf{W} \vec{P}_{\text{ss}} = 0$. The solution is given by
\begin{equation}
\begin{aligned}
&\vec{P}_{\text{ss}}/\mathcal{N} = \\&\begin{pmatrix}
\Gamma_{\ket*{E_{\us \us}} \rightarrow \ket{E_-}} \Gamma_{\ket{E_-} \rightarrow \ket*{E_{\ds \ds}}} \Gamma_{\ket*{E_+} \rightarrow \ket{E_{\us \us}}}+\Gamma_{\ket*{E_{\us \us}} \rightarrow \ket{E_-}} \Gamma_{\ket{E_-} \rightarrow \ket*{E_{\ds \ds}}} \Gamma_{\ket*{E_{+}} \rightarrow \ket*{E_{\ds \ds}}}+\Gamma_{\ket{E_-} \rightarrow \ket*{E_{\ds \ds}}} \Gamma_{\ket*{E_{\us \us}} \rightarrow \ket{E_+}} \Gamma_{\ket*{E_{+}} \rightarrow \ket*{E_{\ds \ds}}}+\Gamma_{\ket*{E_{\us \us}} \rightarrow \ket{E_+}} \Gamma_{\ket*{E_{+}} \rightarrow \ket*{E_{\ds \ds}}} \Gamma_{\ket*{E_-} \rightarrow \ket{E_{\us \us}}} \\ 
\Gamma_{\ket*{E_{\us \us}} \rightarrow \ket{E_-}} \Gamma_{\ket*{E_+} \rightarrow \ket{E_{\us \us}}} \Gamma_{\ket*{E_{\ds \ds}} \rightarrow \ket{E_{+}}}+\Gamma_{\ket*{E_{\us \us}} \rightarrow \ket{E_-}} \Gamma_{\ket*{E_+} \rightarrow \ket{E_{\us \us}}} \Gamma_{\ket*{E_{\ds \ds}} \rightarrow \ket{E_{-}}}+\Gamma_{\ket*{E_{\us \us}} \rightarrow \ket{E_-}} \Gamma_{\ket*{E_{+}} \rightarrow \ket*{E_{\ds \ds}}} \Gamma_{\ket*{E_{\ds \ds}} \rightarrow \ket{E_{-}}}+\Gamma_{\ket*{E_{\us \us}} \rightarrow \ket{E_+}} \Gamma_{\ket*{E_{+}} \rightarrow \ket*{E_{\ds \ds}}} \Gamma_{\ket*{E_{\ds \ds}} \rightarrow \ket{E_{-}}} \\ 
\Gamma_{\ket*{E_{\us \us}} \rightarrow \ket{E_-}} \Gamma_{\ket{E_-} \rightarrow \ket*{E_{\ds \ds}}} \Gamma_{\ket*{E_{\ds \ds}} \rightarrow \ket{E_{+}}}+\Gamma_{\ket{E_-} \rightarrow \ket*{E_{\ds \ds}}} \Gamma_{\ket*{E_{\us \us}} \rightarrow \ket{E_+}} \Gamma_{\ket*{E_{\ds \ds}} \rightarrow \ket{E_{+}}}+\Gamma_{\ket*{E_{\us \us}} \rightarrow \ket{E_+}} \Gamma_{\ket*{E_{\ds \ds}} \rightarrow \ket{E_{+}}} \Gamma_{\ket*{E_-} \rightarrow \ket{E_{\us \us}}}+\Gamma_{\ket*{E_{\us \us}} \rightarrow \ket{E_+}} \Gamma_{\ket*{E_-} \rightarrow \ket{E_{\us \us}}} \Gamma_{\ket*{E_{\ds \ds}} \rightarrow \ket{E_{-}}} \\ 
\Gamma_{\ket{E_-} \rightarrow \ket*{E_{\ds \ds}}} \Gamma_{\ket*{E_+} \rightarrow \ket{E_{\us \us}}} \Gamma_{\ket*{E_{\ds \ds}} \rightarrow \ket{E_{+}}}+\Gamma_{\ket*{E_+} \rightarrow \ket{E_{\us \us}}} \Gamma_{\ket*{E_{\ds \ds}} \rightarrow \ket{E_{+}}} \Gamma_{\ket*{E_-} \rightarrow \ket{E_{\us \us}}}+\Gamma_{\ket*{E_+} \rightarrow \ket{E_{\us \us}}} \Gamma_{\ket*{E_-} \rightarrow \ket{E_{\us \us}}} \Gamma_{\ket*{E_{\ds \ds}} \rightarrow \ket{E_{-}}}+\Gamma_{\ket*{E_{+}} \rightarrow \ket*{E_{\ds \ds}}} \Gamma_{\ket*{E_-} \rightarrow \ket{E_{\us \us}}} \Gamma_{\ket*{E_{\ds \ds}} \rightarrow \ket{E_{-}}} 
\end{pmatrix}.
\end{aligned}
\end{equation}
Here, $\mathcal{N}$ insures conservation of probability, $P_{\text{ss}}(\ket*{E_{\ds \ds}}) + P_{\text{ss}}(\ket*{E_{-}}) + P_{\text{ss}}(\ket*{E_{+}}) + P_{\text{ss}}(\ket*{E_{\us \us}}) = 1$. 

\paragraph{Approximate rates.}

We wish to obtain an approximate solution to the populations, the spin current, and the quantum Fisher information close to the point of interest 
\begin{equation}
J_{x,0} = J_C \left( 1-\frac{h_2}{2J_{23}} \right).
\end{equation}
In the main text, we found that a dramatic change in populations occur for $J_x \sim J_{x,0}$. In order to resolve and explain this, we write the distance from this value as
\begin{equation}
\delta J_x = J_x - J_{x,0},
\end{equation}
and assume $|\delta J_x| \ll J$. In the main text, we studied the the regime $J,h_2 \ll J_{23}$ where all other couplings are of the same order $J_x \sim J_C \sim J \sim \gamma_1$. 
Using these assumptions, we can find an approximate form of the rates. As an example, we can use $\Gamma_{\ket*{E_{\us \us}} \rightarrow \ket{E_-}}$. First, the matrix elements are calculated
\begin{subequations}
\begin{alignat}{1}
M^C_{\ket{E_{\uparrow \uparrow }} \rightarrow \ket{E_-}} &= 2 \left( \bra{\Psi_-} - \frac{h_2}{4J_{23}} \bra{\Psi_+} \right)\left[ J_x \hat{\sigma}_-^{(2)} + J_C \hat{\sigma}_-^{(3)}\right] \ket{E_{\uparrow \uparrow}} \simeq -\sqrt{2} \left( 1+\frac{h_2}{4J_{23}} \right) \delta J_x, \\ 
M_{\ket{E_{\us \us}} \rightarrow \ket{E_-}}^H &= 2J \left( \bra{\Psi_-} - \frac{h_2}{4J_{23}} \bra{\Psi_+} \right)\left[ \hat{\sigma}_-^{(2)} + \hat{\sigma}_-^{(3)}\right] \ket{E_{\uparrow \uparrow}} \simeq -\frac{h_2 J}{\sqrt{2} J_{23}}.
\end{alignat}
\end{subequations}
The rates become
\begin{subequations}
\begin{alignat}{1}
\Gamma^C_{\ket{E_{\uparrow \uparrow }} \rightarrow \ket{E_-}} &= \frac{2 \delta J_x^2 \gamma_1}{2\delta J_x^2 + (2h_1-2J_{23}-h_2)^2 + \gamma_1^2/4} \simeq \frac{2\delta J_x^2 \gamma_1}{(2h_1-2J_{23}-h_2)^2 + \gamma_1^2/4},\\
\Gamma^H_{\ket{E_{\uparrow \uparrow }} \rightarrow \ket{E_-}} &= \frac{\frac{h_2^2 J^2}{2 J_{23}^2} (n+1)\gamma_4}{(2J_{23}+h_2)^2 + \gamma_4^2 (2n+1)^2/4} \simeq \frac{ h_2^2 J^2 (n+1) \gamma_4}{2J_{23}^2 \left(4J_{23}^2 + (2n+1)^2\gamma_4^2 /4\right)} .
\end{alignat}
\end{subequations}
Here, the expression for $E_- - E_{\uparrow \uparrow}$ is only taken to linear order in $h_2$, and we have assumed that $\delta J_x^2 \ll (2h_1 -2J_{23} - h_2)^2/2 +\frac{\gamma_1^2}{8}$. We will see later that $|\delta J_x|/J$ is of order $h_2/J_{23}$, and therefore, this is a valid assumption. We have assumed $h_1 \simeq J_{23}$, but we keep $h_1$ in the expressions such that small deviations from this value are allowed. The contribution from the hot bath is small compared to the contribution from the cold bath, and therefore, it is neglected. This is seen by noting that $|\delta J_x|/J$ is of the order $h_2/J_{23}$ and $J/J_{23}$. The term could also be kept for now and thrown away by direct comparison later. The full list of rates becomes
\begin{subequations}
\begin{alignat}{1}
\Gamma_{\ket*{E_{\us \us}} \rightarrow \ket{E_-}} &\simeq \frac{2 \gamma_1 \delta J_x^2}{ (2h_1 - 2J_{23} - h_2)^2 + \gamma_1^2/4 }, \\
\Gamma_{\ket{E_-} \rightarrow \ket*{E_{\ds \ds}}} &\simeq \frac{ \left(\sqrt{2} \delta J_x - \frac{\sqrt{2} h_2 J_C}{J_{23}}\right)^2 \gamma_1 }{(2h_1+2J_{23})^2} + \frac{ h_2^2 J^2 (n+1) \gamma_4}{2J_{23}^2 \left(4J_{23}^2 + (2n+1)^2\gamma_4^2 /4\right)}, \\
\Gamma_{\ket*{E_{\us \us}} \rightarrow \ket{E_+}} &\simeq \frac{8 J_C^2 \gamma_1}{(2h_1 + 2J_{23})^2} + \frac{ 8J^2 (n+1) \gamma_4}{4J_{23}^2 + (2n+1)^2 \gamma_4^2/4}, \\
\Gamma_{\ket*{E_{+}} \rightarrow \ket*{E_{\ds \ds}}} &\simeq \frac{8 J_C^2 \gamma_1}{8 J_C^2 + (2h_1 -2J_{23} -h_2)^2 + \gamma_1^2/4},   \\
\Gamma_{\ket*{E_{\ds \ds}} \rightarrow \ket{E_{-}}} &\simeq \Gamma_{\ket*{E_-} \rightarrow \ket{E_{\us \us}}} \simeq  \frac{h_2^2 J^2 n \gamma_4 }{2 J_{23}^2\left(4J_{23}^2 + (2n+1)^2\gamma_4^2 /4\right)}, \\
\Gamma_{\ket*{E_{\ds \ds}} \rightarrow \ket{E_{+}}} &\simeq \Gamma_{\ket*{E_+} \rightarrow \ket{E_{\us \us}}} \simeq \frac{ 8J^2 n \gamma_4}{4J_{23}^2 + (2n+1)^2 \gamma_4^2/4}.
\end{alignat}
\end{subequations}

Since the quantum Wheatstone bridge would ideally function the same for all values of $J_x$, the final result should depend only on $\delta J_x$ and not $J_C$ or $J_x$. This is achieved by picking $\gamma_4$ such that the second term in $\Gamma_{\ket{E_-} \rightarrow \ket*{E_{\downarrow \downarrow}}}$ and $\Gamma_{\ket*{E_{\uparrow \uparrow}} \rightarrow \ket*{E_{+}}}$ is larger than the first term. 
For this, we pick $\gamma_1 \ll \gamma_4 \leq 4 J_{23}/(2n+1)$, although $\gamma_4$ could be much larger than the last inequality as long as the assumption on $\Gamma_{\ket{E_-} \rightarrow \ket*{E_{\downarrow \downarrow}}}$ and $\Gamma_{\ket*{E_{\uparrow \uparrow}} \rightarrow \ket*{E_{+}}}$ is fulfilled. 
Finally, we assume that $8J_C^2 \gg (2h_1 - 2J_{23} - h_2)^2 + \gamma_1^2/4$. The new rates are

\begin{subequations}
 \label{rates_app}
\begin{alignat}{1}
\Gamma_{\ket*{E_{\us \us}} \rightarrow \ket{E_-}} &\simeq \frac{2 \gamma_1 \delta J_x^2}{ (2h_1 - 2J_{23} - h_2)^2 + \gamma_1^2/4 }, \\
\Gamma_{\ket{E_-} \rightarrow \ket*{E_{\ds \ds}}} &\simeq \frac{ h_2^2 J^2 (n+1) \gamma_4}{2J_{23}^2 \left(4J_{23}^2 + (2n+1)^2\gamma_4^2 /4\right)},  \\
\Gamma_{\ket*{E_{\us \us}} \rightarrow \ket{E_+}} &\simeq \frac{ 8J^2 (n+1) \gamma_4}{4J_{23}^2 + (2n+1)^2 \gamma_4^2/4}, \\
\Gamma_{\ket*{E_{+}} \rightarrow \ket*{E_{\ds \ds}}} &\simeq \gamma_1 ,\\
\Gamma_{\ket*{E_{\ds \ds}} \rightarrow \ket{E_{-}}} &\simeq \Gamma_{\ket*{E_-} \rightarrow \ket{E_{\us \us}}} \simeq  \frac{h_2^2 J^2 n \gamma_4 }{2 J_{23}^2\left(4J_{23}^2 + (2n+1)^2\gamma_4^2 /4\right)} , \\
\Gamma_{\ket*{E_{\ds \ds}} \rightarrow \ket{E_{+}}} &\simeq \Gamma_{\ket*{E_+} \rightarrow \ket{E_{\us \us}}} \simeq \frac{ 8J^2 n \gamma_4}{4J_{23}^2 + (2n+1)^2 \gamma_4^2/4}.
\end{alignat}
\end{subequations}

\paragraph{Steady state solution, $\vec{P}_{\text{ss}}$.} Solving for the steady state, $\mathbf{W} \vec{P}_{\text{ss}} = 0$, using the the rates from above, we obtain
\begin{equation}
\vec{P}_{\text{ss}} \simeq \frac{1}{(K_1+K_2-1)\delta J_x^2 + \frac{\Lambda^2}{4}}\begin{pmatrix}
(K_1 - 1) \delta J_x^2 + \frac{2n+1}{3n+1} \frac{\Lambda^2}{4} \\
K_2 \delta J_x^2 + \frac{n}{3n+1} \frac{\Lambda^2}{4}  \\
\frac{(n+1) h_2^2   }{16 nJ_{23}^2 } \delta J_x^2 + \frac{8 n(2n+1) J^2  \gamma_4}{(3n+1) \gamma_1 \left(4J_{23}^2 + (2n+1)^2\gamma_4^2 /4\right)} \frac{\Lambda^2}{4} \\
K_3 \frac{n^2 (2n+1)}{(n+1)(3n+1)}\frac{8  J^2 \gamma_4}{\left(4J_{23}^2 + (2n+1)^2\gamma_4^2 /4\right) \gamma_1} \frac{\Lambda^2}{4}
\end{pmatrix}.
\end{equation}
All the populations are of Lorentzian form with full width at half maximum
\begin{equation}
\Lambda = \sqrt{\frac{(n+1)(3n+1)}{2n^2}} \frac{ h_2  \sqrt{\left[2h_1 - 2J_{23} - h_2\right]^2 + \gamma_1^2/4 }  }{2 J_{23} }. \label{FWHM_app}
\end{equation}

Furthermore, we have defined the three constants $K_1$, $K_2$ and $K_3$ to be
\begin{subequations}
\begin{alignat}{1}
K_1 &= 1 + \frac{n+1}{n^2} \frac{  h_2^2 \gamma_1 }{32 J^2 \gamma_4} \frac{4J_{23}^2 + (2n+1)^2\gamma_4^2 /4 }{ 4 J_{23}^2 } \simeq 1 ,\\
K_2 &=  1 + \frac{1}{n} \frac{h_2^2 \gamma_1}{32 J^2 \gamma_4}\frac{4J_{23}^2 + (2n+1)^2\gamma_4^2 /4}{4J_{23}^2} \simeq 1 ,\\
K_3 &= 1 + \frac{1}{2n+1} \frac{h_2^2 \gamma_1 }{32J^2 \gamma_4} \frac{4J_{23}^2 + (2n+1)^2\gamma_4^2 /4 }{4J_{23}^2} \simeq 1,
\end{alignat}
\end{subequations}
which are approximately unity within the approximations. The new steady state is
\begin{align}
\vec{P}_{\text{ss}} \simeq \frac{1}{\delta J_x^2 + \frac{\Lambda^2}{4}}\begin{pmatrix}
 \frac{2n+1}{3n+1} \frac{\Lambda^2}{4} \\
\delta J_x^2 + \frac{n}{3n+1} \frac{\Lambda^2}{4}  \\
\frac{(n+1) h_2^2   }{16 nJ_{23}^2 } \delta J_x^2 + \frac{8 n(2n+1) J^2  \gamma_4}{(3n+1) \gamma_1 \left(4J_{23}^2 + (2n+1)^2\gamma_4^2 /4\right)} \frac{\Lambda^2}{4} \\
 \frac{n^2 (2n+1)}{(n+1)(3n+1)}\frac{8  J^2 \gamma_4}{\left(4J_{23}^2 + (2n+1)^2\gamma_4^2 /4\right) \gamma_1} \frac{\Lambda^2}{4} \label{pop}
\end{pmatrix}.
\end{align}

Since the solution was found to lowest order in $h/J_{23}, J/J_{23}$, and $|\delta J_x|$, other parameters have to be dominant i.e. $n \gg J/J_{23}$. To summarize, the approximations are $J\sim J_C \sim J_x$; $h_2,J \gg J_{23}$; $h_1 \simeq J_{23}$; $\delta J_x^2 \ll (2h_1 - 2J_{23} - h_2)^2 + \gamma_1^2/4 \ll 8J_C^2$; $\gamma_1 \ll \gamma_4 \leq 4 J_{23}/(2n+1)$; $n\gg J/J_{23}$; and $|\delta J_x| \ll J$.

\paragraph{Quantum Fisher information, $\mathcal{F}$.}

Next, we find an expression for the quantum Fisher information, $\mathcal{F}$. The populations $P_{\text{ss}}(\ket*{E_+})$ and $P_{\text{ss}}(\ket*{E_{\us \us}})$ are of order $J^2/J_{23}^2$, and therefore, the Fisher information is predominantly determined by the change in the populations $P_{\text{ss}}(\ket*{E_{\ds \ds}})$ and $P_{\text{ss}}(\ket*{E_-})$. The Fisher information becomes
\begin{equation}
\mathcal{F} \simeq \frac{1}{P_{\text{ss}}(\ket*{E_{\ds \ds}})} \left( \frac{\partial}{\partial J_x} P_{\text{ss}}(\ket*{E_{\ds \ds}}) \right)^2 + \frac{1}{P_{\text{ss}}(\ket*{E_{-}})} \left( \frac{\partial}{\partial J_x} P_{\text{ss}}(\ket*{E_{-}}) \right)^2.
\end{equation}

\begin{figure}[t]
\centering 
\includegraphics[width=1.0 \linewidth, angle=0]{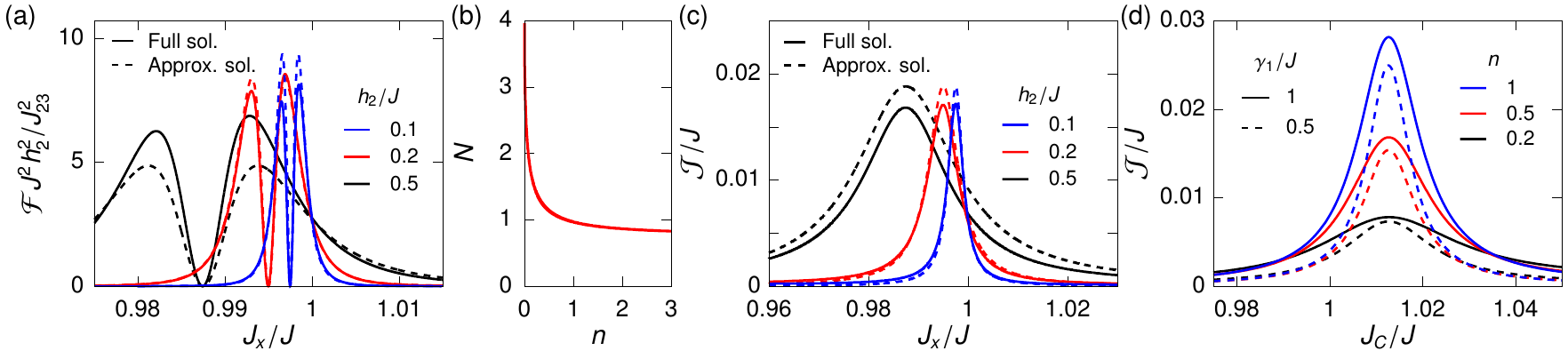}
\caption{(a) Quantum Fisher information, $\mathcal{F}$, and (c) spin current as a function of $J_x$ for different values of $h_2$. The solid line denotes the full theory while the dashed line shows the approximate solutions \eqref{fish} and \eqref{app_cur}.
(b) Value of $N(n)$ from Eq.~\eqref{N} as a function of n. (d) $\mathcal{J}$ as a function of
$J_x$ for different values of the cold bath coupling strength $\gamma_1$ and hot bath
parameter $n$.
}
\label{figure4}
\end{figure}

Putting in the populations from Eq.~\eqref{pop}, we get the approximate expression
\begin{align}
\mathcal{F} \simeq \frac{4 \delta J_x^2 \Lambda^2 \left(2 n + 1 \right)}{\left(\delta J_x^2 + \frac{\Lambda^2}{4}\right)^2 \left(n\Lambda^2+4(3n + 1)\delta J_x^2 \right)}. \label{fish}
\end{align}

This expression is compared to the full solution in Fig.~\ref{figure4}(a). Thus the maximum quantum Fisher information is
\begin{align}
\mathrm{max}(\mathcal{F}) \simeq \frac{4 N(n) }{\Lambda^2},
\end{align}
where
\begin{align}
N(n) = \frac{64 \left(2\,n+1\right)\left(3\,n+1\right) \left(\sqrt{n(25n+8)} - n\right)}{\left(3\,n+\sqrt{n(25n+8)}\right)\,{\left(11\,n+\sqrt{n(25n+8)}+4\right)}^2}. \label{N}
\end{align}

From Fig.~\ref{figure4}(b), we see that $\frac{3}{4} < N(n) \leq 4$ and the scale of the quantum Fisher information is, therefore, predominantly determined by $1/\Lambda^2$.

\paragraph{Spin current, $\mathcal{J}$.}

As mentioned in the main text, the spin current is defined as the number of spin excitations decaying to the cold bath per unit of time. This is equal to the population of the excited state of spin 1, $\mathrm{tr} \{ \hat{\sigma}_+^{(1)} \hat{\sigma}_-^{(1)} \hat{\rho}_{\text{ss}} \}$, times the rate of decay from the excited state, $\gamma_1$. Since spin 1 has been traced away, we instead look at the four transitions $|E_{\us \us} \rangle \rightarrow |E_{+} \rangle$, $|E_{\us \us} \rangle \rightarrow |E_{-} \rangle$, $|E_{+} \rangle \rightarrow |E_{\ds \ds} \rangle$, and $|E_{-} \rangle \rightarrow |E_{\ds \ds} \rangle$ driven by the cold bath. When any of the four transitions occur, a single spin excitation is absorbed by the cold bath. Therefore, the spin current is the sum of the rate for each transition multiplied by the probability of the interface being in the initial state 
\begin{align}
\mathcal{J} = \Gamma_{|E_{\us \us} \rangle \rightarrow |E_{+} \rangle} P_{\text{ss}}(\ket*{E_{\us \us}}) + \Gamma_{|E_{\us \us} \rangle \rightarrow |E_{-} \rangle} P_{\text{ss}}(\ket*{E_{\us \us}}) + \Gamma_{|E_{+} \rangle \rightarrow |E_{\ds \ds} \rangle} P_{\text{ss}}(\ket{E_{+}}) + \Gamma_{|E_{-} \rangle \rightarrow |E_{\ds \ds} \rangle} P_{\text{ss}}(\ket{E_{-}}).
\end{align}
Both the populations $P_{\text{ss}}(\ket*{E_+})$ and $P_{\text{ss}}(\ket*{E_{\us \us}})$ are of order $J^2/J_{23}^2$. From the rates \eqref{rates_app}, it is evident that the spin current going into the cold bath is dominated by the decay of the state $\ket{E_+}$ through the channel $\Gamma_{\ket{E_+}\rightarrow \ket*{E_{\ds \ds}}}$. The spin current can be written
\begin{align}
\mathcal{J} \simeq \gamma_1 P_{\text{ss}}(\ket{E_+}) \simeq \frac{\mathcal{J}_{\infty} \delta J_x^2 + \mathcal{J}_{0} \frac{\Lambda^2}{4}}{\delta J_x^2 + \frac{\Lambda^2}{4}}. \label{app_cur}
\end{align}

$\mathcal{J}_0 = \frac{n (2n+1 )}{ 3n+1} \frac{8 \gamma_4 J^2}{ 4J_{23}^2 + (2n+1)^2\gamma_4^2 /4}$ is the current at $\delta J_x = 0$, and $\mathcal{J}_{\infty} = \frac{(n+1) \gamma_1 h_2^2}{16 n J_{23}^2}$ is the current for $|\delta J_x| \rightarrow \infty$. 
Note that this expression is derived under the assumption of $|\delta J_x| \ll J$ and hence not valid for $|\delta J_x| \rightarrow \infty$, and therefore, $\mathcal{J}_\infty$ should not be taken literally. The full width at half maximum is the same as before, Eq.~\eqref{FWHM_app}.
This value of the current is plotted along side the exact value in Fig.~\ref{figure4}(c). 
Finally, we study the effect of the other parameters on the spin current. The exact current is plotted as a function of $J_C$ for different values of both $\gamma_1$ and $n$ in Fig. \ref{figure4}(d). We observe that larger $\gamma_1$ results in a larger width, $\Lambda$, and a larger $\mathcal{J}_\infty$. Furthermore, we see that larger $n$ results in larger $\mathcal{J}_0$ and slightly larger $\Lambda$. This is generally the behavior expected from the expressions of $\mathcal{J}_0$, $\mathcal{J}_\infty$, and $\Lambda$ from above. Contrary to what we would expect, a larger $\gamma_1$ changes $\mathcal{J}_0$. To capture this behavior, higher orders have to be included in Eq.~\eqref{rates_app}.
We find that, the ratio between the maximum current $\mathcal{J}_0$ and the minimum current $\mathcal{J}_\infty$ is
\begin{align}
\frac{\mathcal{J}_0}{\mathcal{J}_{\infty}} = \frac{128 n^2 (2n+1)}{ (n+1)(3n + 1)} \frac{\gamma_4 J^2 J_{23}^2 }{  h_2^2 \gamma_1 \left(4J_{23}^2 + (2n+1)^2\gamma_4^2 /4\right)}.
\end{align}

This ratio is mainly determined by $J/h_2$ and less sensitive to $J/J_{23}$.

\begin{figure}[t]
\centering 
\includegraphics[width=0.9 \linewidth, angle=0]{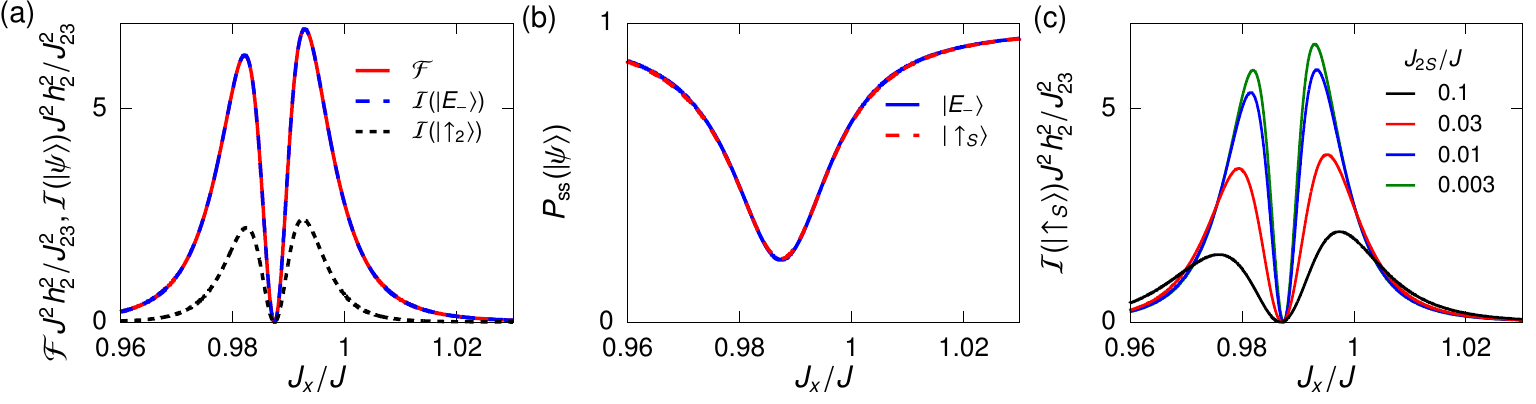}
\caption{(a) Quantum Fisher information and classical Fisher information as a function of $J_x$. (b) Steady-state population of $\ket{E_-}$ and $\ket{\us_S}$ as a function of $J_x$ for $J_{2S} = 0.01J$. (c) Classical Fisher information as a function of $J_x$ for different values of $J_{2S}$. For all plots $J_C = J$ was used.
}
\label{figure6}
\end{figure}

\section{Optimal measurement basis}

To find the optimal measurement basis, we need to define the classical fisher information \cite{doi:10.1142/S0219749909004839, PhysRevLett.106.153603}
\begin{align*}
\mathcal{I} = \sum_a P(a|J_x)\left( \frac{\partial}{\partial J_x} \ln \big(P(a|J_x) \big) \right)^2,
\end{align*}
where $P(a|J_x)$ is the probability of the outcome $a$ of the measurement given $J_x$. Therefore, the classical Fisher information depends on the measurement performed. It is related to the quantum fisher information through the inequality $\mathcal{F} \geq \mathcal{I}$. The optimal measurement is the one for which this is an equality. For a measurement that yields $a=1$ for the state $\ket{\psi}$ and $a=0$ otherwise, the conditional probabilites become
\begin{align}
P(1|J_x) = P(\ket{\psi}) \quad \text{and} \quad P(0|J_x) = 1-P(\ket{\psi}),
\end{align}
where $P(\ket{\psi}) = \mel{\psi}{\hat{\rho}}{\psi}$. The classical Fisher information for this measurement becomes
\begin{align*}
\mathcal{I}(\ket{\psi}) = \frac{1}{P_\text{ss}(\ket{\psi})} \left( \frac{\partial }{\partial J_x} P_\text{ss}(\ket{\psi}) \right)^2 + \frac{1}{1-P_\text{ss}(\ket{\psi})} \left( \frac{\partial }{\partial J_x} \left[1-P_\text{ss}(\ket{\psi})\right] \right)^2.
\end{align*}
Under the assumption $P(\ket{\ds \ds}) + P(\ket{E_-}) \simeq 1$ and a diagonal density matrix and comparing with the previous section, we see that $\mathcal{F} \simeq \mathcal{I}(\ket{E_-}) \simeq \mathcal{I}(\ket{\ds \ds})$. Therefore, the measurement of either $\ket{\ds \ds}$ or $\ket{E_-}$ constitutes an optimal measurement. Both the quantum Fisher information and the classical Fisher information for he state $\ket{E_-}$ is plotted in Fig.~\ref{figure6}(a). $\ket{E_-}$ is an entangled state, and therefore, it can be challenging to measure. In the main text, we proposed measuring spin 2 only. This corresponds to measuring for spin up on spin 2 or $\ket{\us_2}$. In Fig.~\ref{figure6}(a), we see that the classical Fisher information for this measurement is smaller, $\mathcal{I}(\ket{\ds_2}) < \mathcal{I}(\ket{E_-})$. However, measuring locally on spin 2 is still effective and experimentally easier. 

Alternatively, we will provide a scheme for measuring the entangled state $\ket{E_-}$. For this, a shadow spin is added interacting with spin 2
\begin{align*}
\hat{H}' = \hat{H} + J_{2S} \hat{X}_{2S} + \frac{\omega - 2J_{23}}{2} \hat{\sigma}^{(S)}_z 
\end{align*}
where $\hat{H}$ is the quantum Wheatstone bridge Hamiltonian from the main text. $\hat{\sigma}_\alpha^{(S)}$ for $\alpha \in \{x, y, z\}$ are the Pauli matrices for the shadow spin. The Hamiltonian is designed to promote the interaction
\begin{align*}
\ket{E_-} \ket{\ds_S} \leftrightarrow \ket{\ds \ds} \ket{\us_S} \rightarrow \ket{E_-} \ket{\us_S}
\end{align*}
where the second part is due to the original mechanism. In Fig.~\ref{figure6}(b), the steady-state populations is plotted for both the entangled state, $\ket{E_-}$, and the spin 2 spin up state, $\ket{\us_S}$. The two populations clearly overlap, and we have reduced the problem of measuring $\ket{E_-}$ to measuring locally on the shadow spin. The classical Fisher information for a measurement of the shadow spin is shown in Fig.~\ref{figure6}(c). Clearly this scheme is most effective for very small couplings $J_{2S}$.

\begin{figure}[t]
\centering
\includegraphics[width=0.7 \linewidth, angle=0]{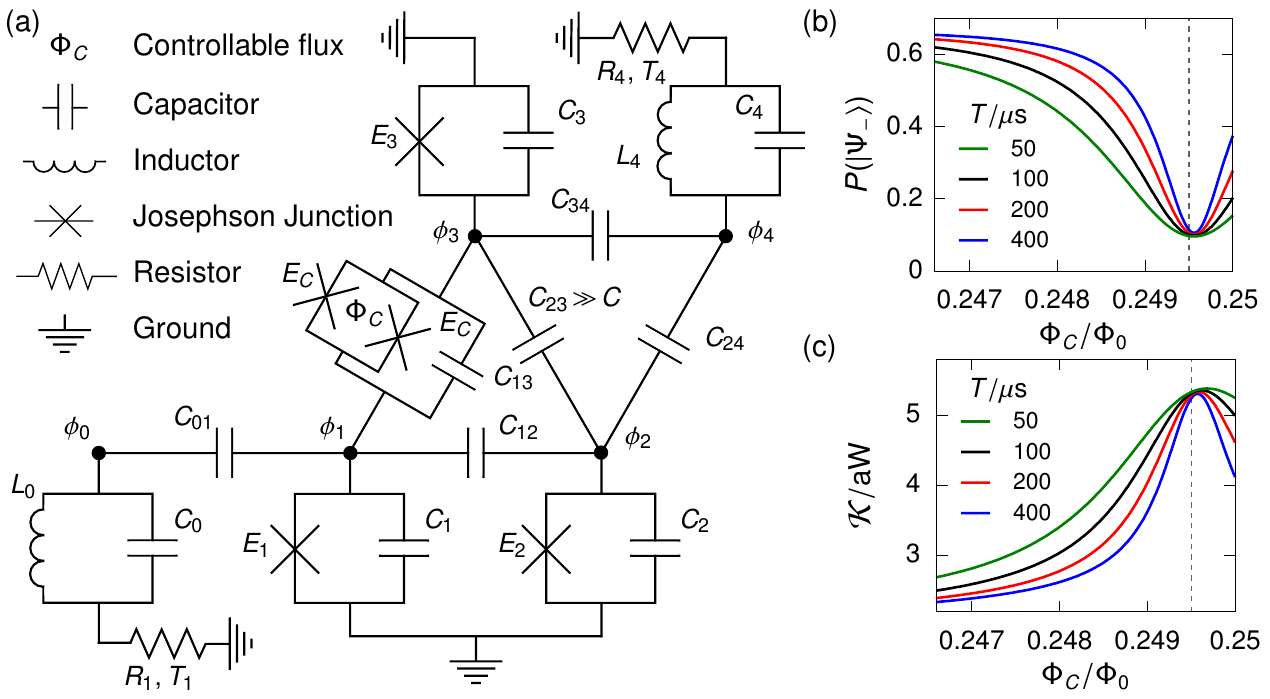}
\caption{(a) Circuit implementation of the quantum Wheatstone bridge. (b) Population $P(\ket{\Psi_-})$ and (c) heat current, $\mathcal{K}$, as a function of the controllable flux, $\Phi_C$, for different coherence times, $T$.
}
\label{figure5}
\end{figure}

\section{Superconducting circuit implementation}
A circuit for implementing the quantum Wheatstone bridge can be seen in Fig.~\ref{figure5}(a). The circuit consists of three transmons implementing qubits 1 to 3 and a microwave resonator coupled to a resistor implementing qubit 4. Qubit 4 is thus replaced by a harmonic oscillator whose correlation functions decay due to the dissipation of the resistor \cite{Ronzani2018}. 
A second resonator is added to the left of qubit 1 in the same manner such that qubit 1 decays. This resonator is not modelled directly to keep the simulations simpler, but it is instead modeled by letting qubit 1 decay. To reach this model, we first write the Lagrangian for the circuit \cite{rasmussen2021superconducting, doi:10.1063/1.5089550, https://doi.org/10.1002/cta.2359}
\begin{equation}
\mathcal{L} 
= \frac{1}{2} \dot{\vec{\Phi}}^T \mathbf{C} \dot{\vec{\Phi}} + \sum_{n=1}^3 E_n \cos \left(2\pi\frac{\Phi_n}{\Phi_0} \right) - \frac{\Phi_4^2}{2L_4} + 2 E_C \left|\cos \left(2\pi \frac{\Phi_C}{\Phi_0}\right)\right| \cos\left(2\pi \frac{\Phi_1 - \Phi_3}{\Phi_0}\right),
\end{equation}
where $\Phi_n$ is the $n$th node flux, $E_n$ is the Josephson energy for the $n$th node, $L_4$ is an inductance, the capacitance matrix is given by
\begin{equation}
\mathbf{C} = \begin{pmatrix}
\tilde{C}_1 & -C_{12} & -C_{13} & 0 \\
-C_{12} & \tilde{C}_2 & - C_{23} & -C_{24} \\
-C_{13} & -C_{23} & \tilde{C}_3 & -C_{34}\\
0 & -C_{24} & -C_{34} & \tilde{C}_4
\end{pmatrix},
\end{equation}
and $\vec{\Phi} = [\Phi_1, \Phi_2, \Phi_3, \Phi_4]^T$. Here, we have defined the effective capacitances and Josephson energies

\begin{subequations}
\begin{alignat}{3}
\tilde{C}_1 &= C_1 + C_{12} + C_{13}, \quad & \tilde{E}_1 &= E_1 + 2E_C \left|\cos \left( 2\pi \frac{\Phi_C}{\Phi_0} \right)\right|, \\
\tilde{C}_2 &= C_2 + C_{12} + C_{23} + C_{24}, \quad &\tilde{E}_2 &= E_2, \\
\tilde{C}_3 &= C_3 + C_{13} + C_{23} + C_{34}, \quad & \tilde{E}_3 &= E_3 + 2E_C \left|\cos \left( 2\pi \frac{\Phi_C}{\Phi_0} \right)\right|,\\
\tilde{C}_4 &= C_4 + C_{24} + C_{34} . \quad &
\end{alignat}
\end{subequations}
Since the circuit consists of coupled transmons, we further assume $C_1, C_2, C_3, C_4 \gg C_{12}, C_{13}, C_{23}, C_{24}, C_{34}$. The conjugate momumtum or the charge is found to be $\vec{q} = \mathbf{C} \vec{\Phi}$ from which the Hamiltonian becomes
\begin{equation}
\mathcal{H} = \frac{1}{2} \vec{q}^T \mathbf{C}^{-1} \vec{q} - \sum_{n=1}^3 E_n \cos \left( 2\pi \frac{\Phi_n}{\Phi_0} \right) + \frac{\Phi_4^2}{2L_4} - 2 E_C \left|\cos \left( 2\pi \frac{\Phi_C}{\Phi_0} \right)\right| \cos\left( 2\pi \frac{\Phi_1 - \Phi_3}{\Phi_0} \right),
\end{equation}
where $\mathbf{C}^{-1}$ is the inverse of the capacitance matrix. Next, we perform the usual quantization. In the transmon regime $\frac{8 \tilde{E}_n}{[\mathbf{C}^{-1}]_{nn}} \frac{\Phi_0^2}{4\pi^2 \hbar^2} \in [60, 100]$, where $[\mathbf{C}^{-1}]_{nn}$ is the $n$th diagonal element of $\mathbf{C}^{-1}$, and $\ev{\frac{2\pi \Phi_n}{\Phi_0}} \ll 1$ for the lowest levels such that the cosine can be expanded to fourth order. This leaves the effective Hamiltonian
\begin{equation}
\begin{aligned}
\hat{H} &= \frac{1}{2} \vec{\hat{q}}^T \mathbf{C}^{-1} \vec{\hat{q}} + \sum_{n=1}^3 \frac{1}{2} \tilde{E}_n  \left( \frac{2 \pi \hat{\Phi}_n}{\Phi_0} \right)^2 - \frac{1}{24} \tilde{E}_n \left( \frac{2 \pi \hat{\Phi}_n}{\Phi_0} \right)^4 + \frac{\hat{\Phi}_4^2}{2L_4}
- 2 E_C \left|\cos \left(\frac{2\pi \Phi_C}{\Phi_0} \right)\right| \frac{2 \pi \hat{\Phi}_1}{\Phi_0} \frac{2 \pi \hat{\Phi}_3}{\Phi_0} \\
&\hspace{1cm} + \frac{1}{12} E_C \left|\cos \left(\frac{2\pi \Phi_C}{\Phi_0} \right)\right| \left[ 2\left(\frac{2 \pi \hat{\Phi}_1}{\Phi_0} \right)^3 \frac{2 \pi \hat{\Phi}_3}{\Phi_0} - 3\left(\frac{2 \pi \hat{\Phi}_1}{\Phi_0} \right)^2 \left(\frac{2 \pi \hat{\Phi}_3}{\Phi_0} \right)^2 + 2\left(\frac{2 \pi \hat{\Phi}_1}{\Phi_0} \right) \left(\frac{2 \pi \hat{\Phi}_3}{\Phi_0} \right)^3 \right].
\end{aligned}
\end{equation}
Next we define the harmonic oscillator ladder operators through

\begin{subequations}
\begin{alignat}{3}
\hat{\Phi}_1 &= \frac{1}{\sqrt{2}} \left( \frac{4\pi^2  \tilde{E}_1 }{\Phi_0^2 \hbar^2 [\mathbf{C}^{-1}]_{11}} \right)^{-1/4} \left( \hat{a}_1^\dag + \hat{a}_1 \right), \quad \quad \quad & \hat{q}_1 &= \frac{i}{\sqrt{2}} \left( \frac{4\pi^2 \hbar^2 \tilde{E}_1 }{\Phi_0^2 [\mathbf{C}^{-1}]_{11}} \right)^{1/4} \left( \hat{a}_1^\dag - \hat{a}_1 \right) , \\
\hat{\Phi}_2 &= \frac{1}{\sqrt{2}} \left( \frac{4\pi^2 \tilde{E}_2 }{\Phi_0^2 \hbar^2 [\mathbf{C}^{-1}]_{22}} \right)^{-1/4} \left( \hat{a}_2^\dag + \hat{a}_2 \right), \quad & \hat{q}_2 &= \frac{i}{\sqrt{2}} \left( \frac{4\pi^2\hbar^2 \tilde{E}_2 }{\Phi_0^2 [\mathbf{C}^{-1}]_{22}} \right)^{1/4} \left( \hat{a}_2^\dag - \hat{a}_2 \right) ,\\
\hat{\Phi}_3 &= \frac{1}{\sqrt{2}} \left( \frac{4\pi^2 \tilde{E}_3 }{\Phi_0^2 \hbar^2 [\mathbf{C}^{-1}]_{33}} \right)^{-1/4} \left( \hat{a}_3^\dag + \hat{a}_3 \right), \quad & \hat{q}_3 &= \frac{i}{\sqrt{2}} \left( \frac{4\pi^2\hbar^2 \tilde{E}_3 }{\Phi_0^2 [\mathbf{C}^{-1}]_{33}} \right)^{1/4} \left( \hat{a}_3^\dag - \hat{a}_3 \right), \\
\hat{\Phi}_4 &= \frac{1}{\sqrt{2}} \left( \frac{1}{\hbar^2 L_4 [\mathbf{C}^{-1}]_{44}} \right)^{-1/4} \left( \hat{a}_4^\dag + \hat{a}_4 \right), \quad & \hat{q}_4 &= \frac{i}{\sqrt{2}} \left( \frac{\hbar^2}{L_4 [\mathbf{C}^{-1}]_{44}} \right)^{1/4} \left( \hat{a}_4^\dag - \hat{a}_4 \right),
\end{alignat}
\end{subequations}


where we have put $\hbar$ back into the expression explicitly. We further simplify the parameters and use the following physically realistic parameters \cite{rasmussen2021superconducting, doi:10.1063/1.5089550, https://doi.org/10.1002/cta.2359}
\begin{equation}
\begin{aligned}
E_2 = E_3 = E = 2\pi \hbar \cdot 20\mathrm{GHz}, \quad E_1 = E + \Delta E = 2\pi \hbar \cdot 24\mathrm{GHz}, \quad L_4 = \frac{\Phi_0^2}{4\pi^2 E} \simeq 8.2\mathrm{nH}, \\ \tilde{C} = \tilde{C}_1 = \tilde{C}_2 = \tilde{C}_3 = \tilde{C}_4 = 80\mathrm{fF}, \quad
C_{12} = C_{24} = C_{34} = 1\mathrm{fF}, \quad C_{13} = 0.95\mathrm{fF}, \quad C_{23} = 7\mathrm{fF}, \quad E_C = 2\pi \hbar \cdot 2\mathrm{GHz}. \hspace{-1.2cm}
\end{aligned}
\end{equation}

In general, the qubit frequencies, $\omega_n$, and exchange couplings, $J_{nm}$, depend on $\Phi_C$. If we assume $\tilde{E}_n \simeq E$ and exploit that $\mathbf{C}$ is almost diagonal, they can be determined approximately through
\begin{subequations}
\begin{alignat}{1}
\hbar \omega_1 &= \frac{2\pi \hbar}{\Phi_0} \sqrt{\frac{E + \Delta E}{\tilde{C}}} - \frac{\pi^2 \hbar^2}{\Phi_0^2 \tilde{C}} \simeq 2\pi\hbar\cdot 6.3 \mathrm{GHz}, \\
\hbar \omega_2 &= \hbar \omega_3 = \frac{2\pi \hbar}{\Phi_0} \sqrt{\frac{E}{\tilde{C}}} - \frac{\pi^2 \hbar^2}{\Phi_0^2 \tilde{C}} \simeq 2\pi\hbar\cdot 5.7 \mathrm{GHz}, \label{omega23}\\
\hbar \omega_4 &= \frac{\hbar}{\sqrt{\mathcal{C}_4 L_4}} \simeq 2\pi\hbar \cdot 6.2 \mathrm{GHz},\\
J_{12} &= \frac{\pi \hbar}{\Phi_0} \frac{C_{12}}{\tilde{C}_1 \tilde{C}_2} (\tilde{C}_1 E_1 \tilde{C}_2 E_2)^{1/4} \simeq 2\pi \hbar \cdot 39\mathrm{MHz},\\
J_{23} &= \frac{\pi \hbar}{ \Phi_0} \frac{C_{23}}{\tilde{C}_1 \tilde{C}_3} (\tilde{C}_2 E_2 \tilde{C}_3 E_3)^{1/4} \simeq 2\pi \hbar \cdot 270\mathrm{MHz}.
\end{alignat}
\end{subequations}
$J_{24}$ and $J_{34}$ can be determined in a similar way, while $J_{13}$ depends on $\Phi_C$ even to lowest order.
The two resistors are modeled through the Lindblad equation
\begin{equation}
\frac{d \hat{\rho}}{d t} = -\frac{i}{\hbar} [\hat{H}, \hat{\rho}] + \mathcal{D}_1[\hat{\rho}] + \mathcal{D}_2[\hat{\rho}] + \mathcal{D}_3[\hat{\rho}] + \mathcal{D}_4[\hat{\rho}].
\end{equation}

The Markovian baths are modeled using the non-unitary parts
\begin{subequations}
\begin{alignat}{1}
\mathcal{D}_{1}[\hat{\rho}] &= \gamma_1  \left( \hat{a}_{1} \hat{\rho} \hat{a}^\dagger_{1} - \frac{1}{2} \{ \hat{a}_1^\dagger \hat{a}_1, \rho \} \right), \\
\mathcal{D}_{2}[\hat{\rho}] &= \frac{1}{T} \left( \hat{a}_{2} \hat{\rho} \hat{a}^\dagger_{2} - \frac{1}{2} \{ \hat{a}_2^\dagger \hat{a}_2, \rho \} \right) + \frac{1}{T} \left( \hat{a}_{2}^\dagger \hat{a}_{2} \hat{\rho} \hat{a}_{2}^\dagger \hat{a}_{2} - \frac{1}{2} \{ \hat{a}_2^\dagger \hat{a}_2 \hat{a}_2^\dagger \hat{a}_2, \rho \} \right), \\
\mathcal{D}_{3}[\hat{\rho}] &= \frac{1}{T} \left( \hat{a}_{3} \hat{\rho} \hat{a}^\dagger_{3} - \frac{1}{2} \{ \hat{a}_3^\dagger \hat{a}_3, \rho \} \right) + \frac{1}{T} \left( \hat{a}_{3}^\dagger \hat{a}_{3} \hat{\rho} \hat{a}_{3}^\dagger \hat{a}_{3} - \frac{1}{2} \{ \hat{a}_3^\dagger \hat{a}_3 \hat{a}_3^\dagger \hat{a}_3, \rho \} \right), \\
\mathcal{D}_{4}[\hat{\rho}] &= \gamma_4 (n + 1) \left( \hat{a}_{4} \hat{\rho} \hat{a}^\dagger_{4} - \frac{1}{2} \{ \hat{a}_4^\dagger \hat{a}_4, \rho \} \right) + \gamma_4 n \left( \hat{a}^\dagger_{4} \hat{\rho} \hat{a}_{4} -\frac{1}{2} \{ \hat{a}_{4} \hat{a}^\dagger_{4}, \rho \} \right).
\end{alignat}
\end{subequations}

The coherence time for qubit 2 and 3 is denoted $T=T_1=T_2$. We have thus assumed $T_1 = T_2$ for simplicity although this is not necessary for the functionality. Decoherence of qubit 1 and resonator 4 is not included explicitly since they already decay due to the two baths. Furthermore, we have assumed the temperature of the left resistor to be small, $k_B T_1 \ll \hbar/\sqrt{C_0 L_0}$. The expression for $\gamma_1$ and $\gamma_4$ depends on the exact configuration of the circuit so we will set
\begin{equation}
\gamma_1 = 2\pi \cdot 10\mathrm{MHz}, \quad \gamma_4 = 2\pi \cdot 500\mathrm{MHz}, \quad n = 0.3\,.
\end{equation}
In Fig.~\ref{figure5}(b), we show the population of the entangled state $P_{\text{ss}}(\ket{\Psi_-}) = \mel{\Psi_-}{\hat{\rho}_{\text{ss}}}{\Psi_-}$ as a function of the external flux $\Phi_C$ for different values of the coherence time $T$. 
In Fig.~\ref{figure5}(c), the heat current, $\mathcal{K} = \omega_1 \gamma_1 \mathrm{tr}\{\hat{a}_1^\dagger \hat{a}_1 \hat{\rho}_{\text{ss}}\}$, exchanged between the cold bath and the system is plotted as a function of $\Phi_C$. 
As a first approximation of the local minimum in $P_{\text{ss}}(\ket{\Psi_-}) = \text{tr}\{ \op{\Psi_-} \hat{\rho}_{\text{ss}} \} $, we can set the hopping between spins 1 and 3 equal the hopping between spins 1 and 2 corresponding to $J_C = J_x$ in the main article. Keeping the lowest two terms when inverting $\mathbf{C}$ the corresponding $\Phi_C$ becomes
\begin{equation}
\frac{\Phi_{C,0}}{\Phi_0} = \frac{1}{2\pi} \acos \left\{ \frac{\sqrt{E (E + \Delta E)}}{2E_C \tilde{C}^2} \Big( C_{13} - C_{12} \Big) \left( \tilde{C} - C_{23}\right) \right\} \simeq 0.24946955.
\end{equation}

This is plotted as a dotted line in Figs~\ref{figure5}(b)-(c). This assumes $\omega_2 = \omega_3$, which we assumed above. However, the Josephson junction $E_C$ changes the frequency of qubit 3 which explains why $\Phi_{C,0}$ is not exact. For the heat current, the critical value of the external flux from above is a better prediction for longer coherence times, $T$. We clearly see that even for experimentally realistic coherence times of $T = 50 \mathrm{\mu s}-100\mathrm{\mu s}$ the mechanism works \cite{doi:10.1146/annurev-conmatphys-031119-050605}.

\end{document}